\documentstyle[emulateapj,multicol,psfig]{article}       
\def\be{\begin{equation}}
\def\ee{\end{equation}}

\def\lsim{\lower 2pt \hbox{$\, \buildrel {\scriptstyle <}\over
         {\scriptstyle \sim}\,$}}

\begin{document}
\newcommand{\figureout}[4]{\psfig{figure=#1,width=#2,angle=#3} 
   \figcaption{#4} }

\title{Off-Beam Gamma-Ray Pulsars and Unidentified \\ 
EGRET Sources in the Gould Belt}

\author{Alice K. Harding\altaffilmark{1} and Bing Zhang\altaffilmark{1,2,3}}
\altaffiltext{1} {Laboratory of High Energy Astrophysics, NASA/Goddard Space  
 Flight Center, Greenbelt, MD 20771}
\altaffiltext{2} {National Research Council Research Associate Fellow} 
\altaffiltext{3} {Current address: Astronomy \& Astrophysics Department, 
Pennsylvania State University, University Park, PA 16802}

\begin{abstract}

We investigate whether gamma-ray pulsars viewed at a large angle to 
the neutron star magnetic pole could contribute to the new population of 
galactic unidentified EGRET sources associated with the Gould Belt.  
The faint, soft nature of these sources is distinctly different from
both the properties of unidentified EGRET sources along the galactic 
plane and of the known gamma-ray pulsars.  We explore the possibility, within
the polar cap model, that some of these sources are emission from
pulsars seen at lines of sight that miss both the bright gamma-ray
cone beams and the radio beam.  The off-beam gamma-rays come from
high-altitude curvature emission of primary particles, are radiated over
a large solid angle and have a much softer spectrum than that of the main
beams.  We estimate that the detectability of such off-beam 
emission is about a factor of 4-5 higher than that of the on-beam 
emission.  At least some of the radio-quiet Gould Belt sources detected
by EGRET could therefore be such off-beam gamma-ray pulsars. GLAST
should be able to detect pulsations in most of these sources.

\end{abstract}

\keywords{stars: neutron; pulsar: general; gamma rays: pulsars}


\section{INTRODUCTION}

The origin of the 170 unidentified gamma-ray point sources detected by the Energetic 
Gamma-Ray Experiment Telescope (EGRET) on board the Compton Gamma-Ray Observatory (CGRO)
throughout its mission (Hartman et al. 1999) is one of the most intriguing questions
of high-energy astrophysics.  Grenier \& Perrot (1999) found that some of the unidentified 
sources are significantly correlated with the Gould Belt of massive stars, a nearby
galactic structure surrounding the Sun.  The Gould Belt is an expanding disk of gas 
and young stars, most with ages less than 30
million years, inclined about $20^0$ to the galactic plane.  
Recently, Gehrels et al. (2000) studied the 
spatial and flux distribution of the 120 steady sources in the third EGRET catalog and 
found that they divide into two groups: higher flux sources distributed along the 
galactic plane with latitudes less than $5^0$, and a new population of lower flux sources 
at mid-latitude
which indeed seem to be associated with the Gould Belt, at a relatively nearby distance 
of 100-300 pc.  The two groups also show significant differences in their spectra,
with the low-latitude sources having average photon spectral index of $2.18 \pm 0.04$
and the mid-latitude sources having average photon index of $2.49 \pm 0.04$.  
None of these sources have known counterparts at other wavelengths.
Grenier \& Perrot (1999) suggested that the Gould Belt $\gamma$-ray sources could 
be young pulsars, formed in supernova explosions of massive stars of the belt.  
However, the relatively soft spectra and luminosities of these sources, averaging
$\sim 6 \times 10^{30}\,{\rm erg\,s^{-1}}$, are not
characteristic of the known $\gamma$-ray pulsars detected by EGRET, which have spectral indices of $1.5 - 2.0$, luminosities $10^{32} - 10^{34}\,{\rm erg\,s^{-1}}$ 
and 0.5-4 kpc distances.

Here, we show that the bulk of the Gould-Belt $\gamma$-ray sources could be pulsars
if their emission is seen at large angles to their magnetic axes, such that we are
missing the bright, hard $\gamma$-ray beams but detecting only the off-beam emission.
According to the polar cap model (e.g. Daugherty \& Harding
1996 [DH96]), $\gamma$-ray emission occurs throughout the entire pulse phase.  
A synchrotron-pair
cascade at low altitude radiates the hard on-beam emission in a hollow cone centered 
on the magnetic pole to produce the bright double-peaked pulses.  Primary electrons
continue to radiate curvature emission on open field lines to high altitudes beyond 
the cascade region, producing a lower level of softer off-beam emission.  Due to the 
flaring of the dipole field lines, this emission may be seen over a large solid angle, 
far exceeding that of the main beams.  Since the radio emission is expected to originate
within ten stellar radii of the neutron star surface, it is quite probable to see
off-beam $\gamma$-ray emission and miss the radio beam.  Such off-beam or off-pulse 
emission may have been detected by EGRET in the Crab, Vela and Geminga $\gamma$-ray 
pulsars during non-pulse phases (Fierro et al. 1998).  We show that
this expected off-beam emission has the right characteristics (spectral index, 
luminosity
and numbers) to account for the EGRET sources associated with the Gould Belt. 

\section{Off-Beam Pulsar Emission}

\subsection{Geometry}

The basic geometry of polar cap emission, projected onto the stellar surface, 
is shown in Figure 1.  The main beam of the $\gamma$-ray cascade emission zone is 
located along the edge of the hollow cone with opening angle $\theta_{\gamma}$, 
centered on the magnetic axis, denoted
by the dotted circle labeled $\gamma$.  The altitude of this zone is about 1 - 2 stellar
radii above the stellar surface (DH96, Harding \& Muslimov 1998).  The off-beam 
$\gamma$-ray emission zone is
located at altitudes above the cascade emission zone, from several stellar radii to the
light cylinder.  Off-beam emission visible to EGRET above 20 MeV is shown as a shaded 
region.  The radio emission can
have a core component, thought to be located at low altitude along the magnetic axis,
and/or a conal component(s), thought to be located in a hollow cone at altitudes of
tens of stellar radii (e.g. Rankin 1993).  In the polar cap model, the radio
emission originates from the secondary pair cascade. In Fig. 1, the possible location
of radio conal emission with opening angle $\theta_R$ is shown as a dashed circle 
labeled R.  The circles centered on the rotation axis, labeled $\zeta_1$ and $\zeta_2$, 
denote the sweep in pulse phase of different observer lines-of-sight viewing on-beam
and off-beam $\gamma$-ray emission respectively.

\hskip -1.0 truecm
\figureout{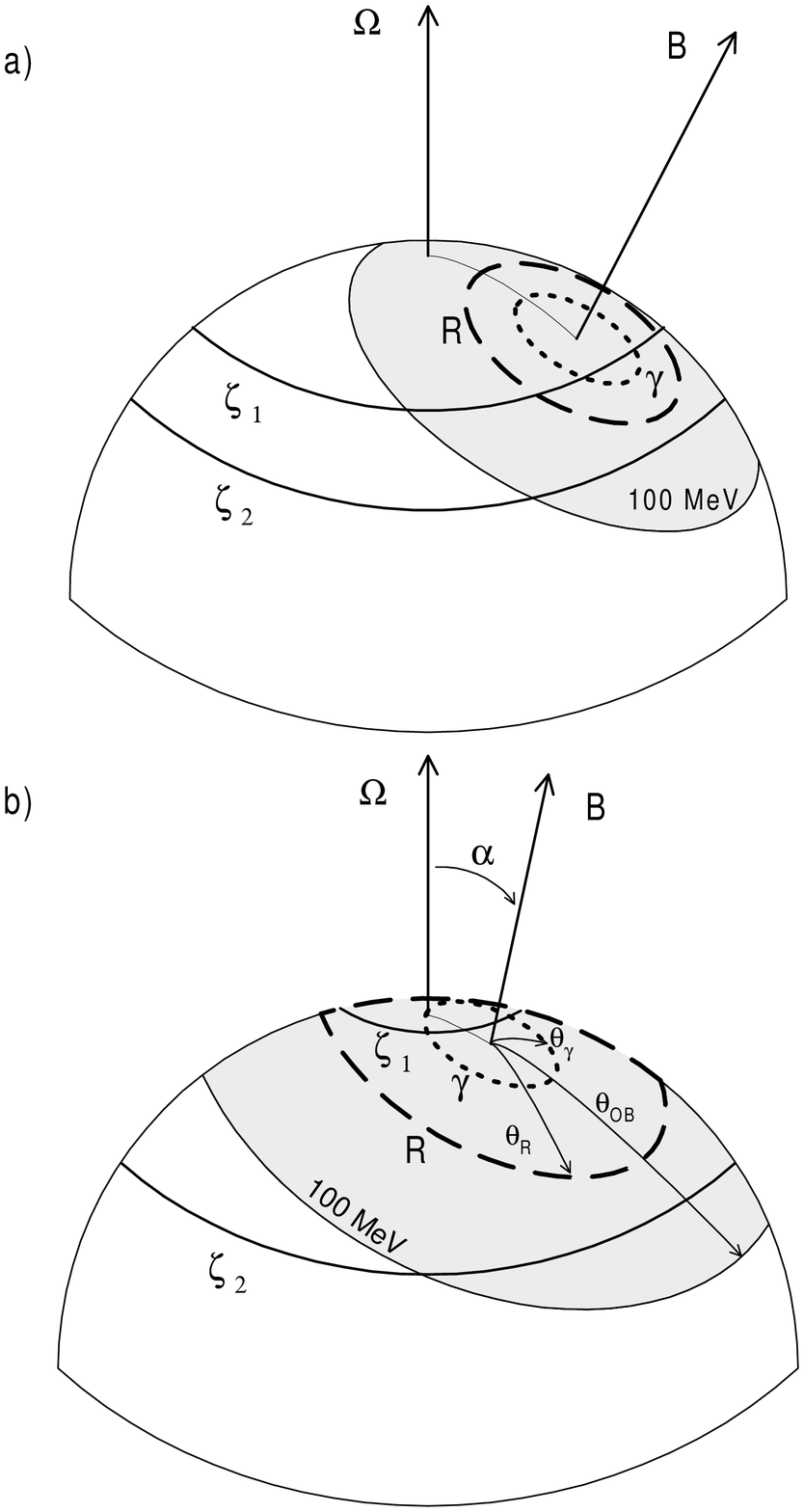}{4.4in}{0}
{The geometry of the polar cap model projected onto
the stellar surface. The dotted circle labeled $\gamma$ is the main 
gamma-ray conal beam, the dashed circle is a conal component labeled
R of the radio emission
beam, and the circle enclosing the shaded region denotes the cone where the typical
curvature radiation emission of the primary particles is 100 MeV.
The shaded area is the above-100 MeV-emission area of the pulsar.
The inclination angle between $\Omega$ and ${\bf B}$ is $\alpha$ and
$\zeta$'s denote different lines of sight. The 
line-of-sight $\zeta_2$ cuts off the main gamma-ray beam and 
misses the radio emission beam, and we identify such a
configurations as the off-beam pulsar candidates for the ``Gould Belt'' faint, soft,
unidentified EGRET sources. (a) $\alpha$ not small: $\zeta_1$ 
gives a typical Vela-type emission configuration; (b) $\alpha$
small: $\zeta_1$ gives a typical Geminga-type emission configuration.
}
\vskip 0.2 truecm

The off-beam gamma-ray emission is mainly produced by curvature radiation of the primary 
particles, well beyond the acceleration and cascade zone. For a certain inclination 
angle $\alpha$ and line of sight
$\zeta$, the off-beam emission zone, between $r_{\rm OB}({\rm min})$ and 
$r_{\rm OB}({\rm max})$, (or the corresponding $\theta_{\rm OB}
({\rm min})$ and $\theta_{\rm OB}({\rm max})$) may be determined by geometry and 
the curvature radiation properties. 
The minimum height, $r_{\rm OB}({\rm min})$, is the lowest height visible to a given
line-of-sight and is defined by the radius at which the tangent to the last open field line, which is ${3\over 2} \theta$ for a
dipole, where $\theta$ is the polar angle, points in the direction
of the observer (which is also the outer edge of the off-beam $\gamma$-ray emission 
cone at that radius):
\be  \label{robmin}
r_{_{\rm OB}}({\rm min}) \simeq {c\over \Omega}\left[{2\over 3}
(\zeta-\alpha)\right]^2
\ee
For $P=0.3$ s, and $\zeta-\alpha = 10^{\rm o}$, we have 
$r_{_{\rm OB}}({\rm min}) \sim 19 R$.

The maximum height at which EGRET-band emission ($> 100$ MeV) can be produced may
be defined as the radius $r_{\rm OB} ({\rm max})$ along the last open field line where
the curvature radiation photon energy $E_0 = 100$ MeV.  
The typical curvature photon energy is $E_0=(3/2)(\gamma^3
\hbar c/\rho)=322 {\rm MeV}\, \gamma_7^3 P^{-1/2}(r/R)^{-1/2}$ for dipolar field 
configuration, where $\gamma_7 \equiv \gamma/10^7$ is the primary particle Lorentz 
factor, $\rho$ is the magnetic field radius of curvature, $P$ is
the pulsar period and $r$ and $R$ are the radii of the emission point and stellar 
radius, respectively.  To be detected by EGRET, one thus requires 
\be   \label{gamma20}
\gamma_7 > \gamma_7({\rm 100~MeV})\equiv \gamma_{100}= 0.68\, P^{1/6}(r/R)^{1/6}. 
\ee

The expression for the Lorentz factor of a particle traveling
along an open dipole field line and subject only to curvature radiation losses is
(Harding 1981)
\begin{equation}  \label{gammar}
{\gamma \over \gamma_0} = \left[1 + {9\over 8}r_e\theta^2 {\gamma_0^3 \over r}
\ln({r\over R_0})\right]^{-1/3},
\end{equation}
where $r_e = e^2/mc^2$ is the classical electron radius and $\gamma_0$ is the 
initial Lorentz factor at $R_0$, the radius of the acceleration zone. 
Using Eqn (\ref{gammar}), Eqn (\ref{gamma20}), and the expression for the last 
open dipole field line,
$\theta^2 = (r\Omega/c)$, we have the following equation for $r_{\rm OB} ({\rm max})$,
\be   \label{robmax}
\left[{\gamma_{100}(r_{_{\rm OB}} ({\rm max})) \over \gamma_{0,7}}\right]^{-3} = 1 + 
{9\over 8}
{r_e\over c}\Omega\gamma_0^3\ln({r_{_{\rm OB}} ({\rm max})\over R_0})
\ee
where $\gamma_{0,7} \equiv \gamma_0/10^7$.
Finding the root of this equation yields values of $r_{\rm OB} ({\rm max})$ for
different values of $\Omega = 2\pi/P$, $\gamma_0$ and $R_0$, which will be limited by the light 
cylinder distance, $\sim r_{LC}/\sin\alpha$ in short-period pulsars.  For
$P = 0.3$ s, $\gamma_0 = 2 \times 10^7$ and $R_0 = 2R$, one finds $r_{\rm OB} ({\rm max})
= 48R$.

\subsection{Relative Observability}

The relative opening angle, $\theta_{_{\rm OB}} \sim (\Omega 
r_{_{\rm OB}}({\rm max})/c)^{1/2}$, 
of the ``100 MeV cone'' with respect to the on-beam $\gamma$-ray cone, 
$\theta_{\gamma} \sim (\Omega r_{\gamma}/c)^{1/2}$, 
is: ${\theta_{_{\rm OB}}/\theta_{\gamma}}\simeq
\left({r_{_{\rm OB}}({\rm max})/r_{\gamma}}\right)^{1/2}$.
For a fixed $\alpha$, the relative detectability of the off-beam 
pulsars and the gamma-radio plus Geminga-like pulsars is (note that only 
geometric effects, but no luminosity selection effect is taken into account): 
\begin{equation}
D(\alpha)=\frac{\cos({\rm Max}[0,\alpha-\theta_{_{\rm OB}}])-\cos({\rm 
Min}[\pi/2,\alpha+\theta_{_{\rm OB}}])}{\cos({\rm Max}[0,\alpha-
\theta_{\gamma}])-\cos({\rm Min}[\pi/2,\alpha+\theta_{\gamma}])}-1
\end{equation}
Assuming a random distribution of $\alpha$, the relative
detectability of the off-beam pulsars and the gamma-radio plus 
Geminga-like pulsars is (Emmering \& Chevalier 1989): 
\begin{equation} \label{D}
D=\frac{(1-\cos\theta_{_{\rm OB}})+(\pi/2-\theta_{_{\rm OB}})\sin\theta_
{_{\rm OB}}}
{(1-\cos\theta_{\gamma})+(\pi/2-\theta_{\gamma})\sin\theta_{\gamma}}-1.
\end{equation}
Given typical values $P=0.3$ s, $r_{\gamma}=2$, and assuming emission along
the ``last open field line'', one gets
$\theta_{_{\rm OB}}/ \theta_{\gamma}\sim 5$ and $D\sim 4$
according to eq.[\ref{D}].  Figure 2 shows the dependence of both $r_{_{\rm OB}}(max)$
and $D$ on pulsar period.
If we regard Geminga and PSR 0656+14 (the strongest gamma
ray pulsar candidate) as the on-beam gamma-ray pulsars in the
Gould Belt, we expect that there might about 8 off-beam gamma-ray
pulsars, a sizable portion of the ~20-40 Gould Belt
unidentified EGRET sources (Grenier et al. 2000).
The luminosity selection effects will modify this number, but 
the above estimate suggests that at least some of the 
Gould Belt sources could be off-beam $\gamma$-ray pulsars.

\vskip -0.5 truecm
\hskip -1.0 truecm
\figureout{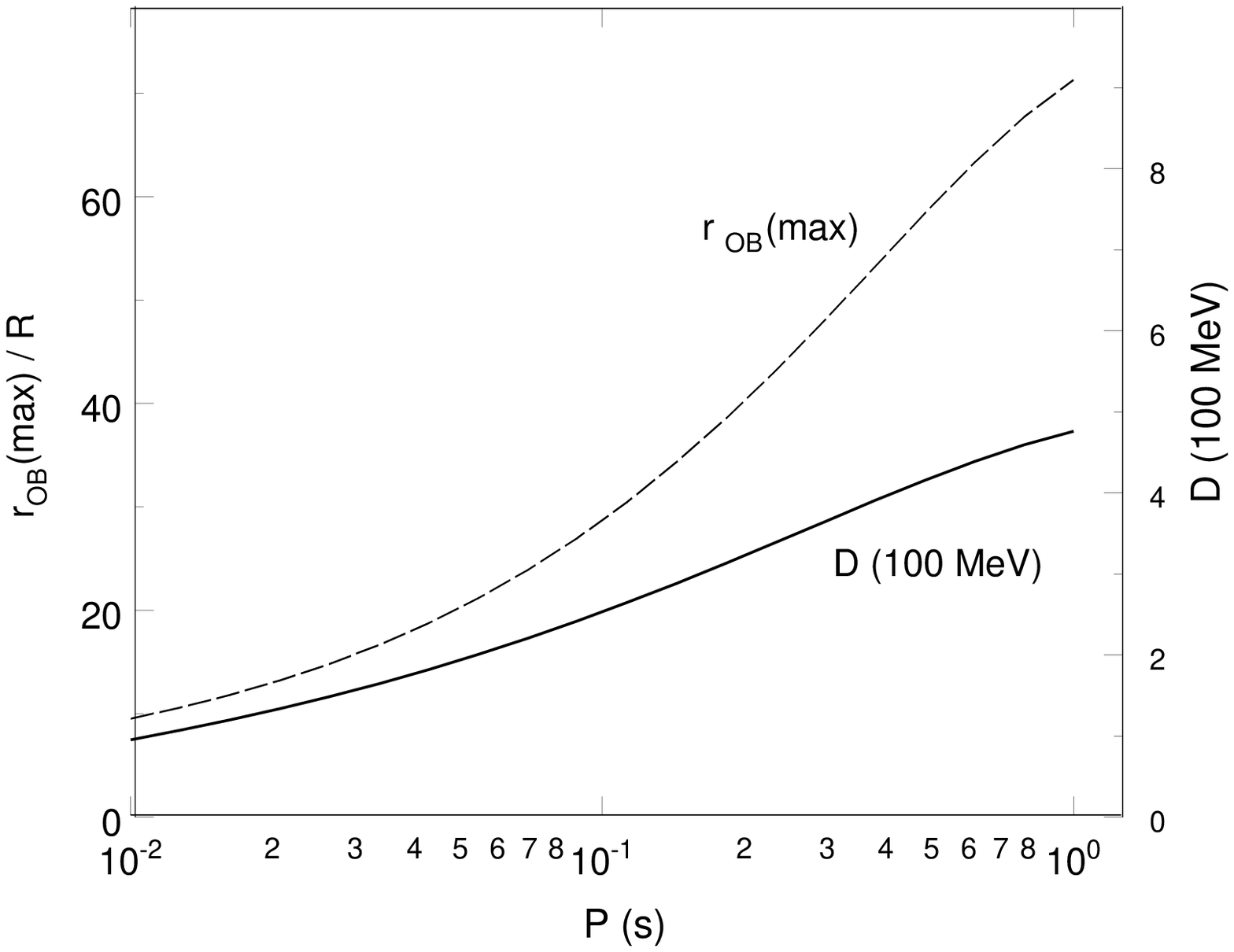}{4.0in}{0}
{Solutions for the maximum radius of off-beam emission, $r_{_{\rm OB}}(\rm 
max)$,
from Eqn. (\ref{robmax}) and for the detectability of off-beam relative to 
on-beam emission, $D$, Eqn (\ref{D}), as a function of period, $P$.}
\vskip 0.5 truecm

Figure 1a and 1b show two different possibilities for observer
orientation relative to the spin and magnetic axes.  In Fig. 1a,
$\zeta_1 > \alpha$ so that the line-of-sight cuts outside magnetic axes,
seeing both bright rims of the $\gamma$-ray cone (and thus a double-peaked
$\gamma$-ray profile) and the radio cone.  
Assuming the trailing radio component is missing, the 
broad band pulse profiles of the Vela pulsar and PSR B1046-58, 
with $\gamma$-ray pulse phase separation $W=0.4$ and phase difference between
the radio and leading $\gamma$-ray pulses of $\delta\phi=0.12$, are consistent 
with this picture over a large $\alpha-\zeta$ phase space as 
long as $\alpha$ is not too large.  In Fig. 1b, $\zeta_1 < \alpha$ so 
that the observer cuts between the spin and magnetic axes, seeing 
both bright rims of the $\gamma$-ray cone (and thus a double-peaked
$\gamma$-ray profile), but missing or grazing the radio cone.  Such a
configuration might describe Geminga-like pulsars.  Thus, the phase space
${\rm Max}(0, \alpha-\theta_{\gamma}) <\zeta < {\rm Min}
(\pi/2, \alpha +\theta_{\gamma})$ comprises radio-loud $\gamma$-ray pulsars 
and Geminga-like pulsars, both with double-peaked $\gamma$-ray profiles.
The phase space $\alpha +\theta_{\gamma} < \zeta < \pi/2$ or
$0 < \zeta < \alpha -\theta_{\gamma}$ contains the off-beam 
$\gamma$-ray pulsars having single peaked $\gamma$-ray profiles.
Thus, we expect that the majority of the $\gamma$-ray pulsars detected
in the Gould Belt will be radio quiet.

\subsection{Luminosity and Spectrum}

Since the energy loss rate for curvature radiation, $\dot\gamma$, 
\be \label{Loff}
\dot\gamma={2\over 3}{\gamma^4
e^2 \over mc^5}\left({c^2 \over \rho}\right)^2 = 6.6\times
10^9 \gamma_7^4 P^{-1}(r/R)^{-1},
\ee
declines with height, the
luminosities of the off-beam sources then mainly depend on 
$\dot\gamma[r_{_{\rm OB}}({\rm min})]$.  A rough estimate of the relative 
luminosities of the on-beam and off-beam $\gamma$-ray emission is then
\begin{equation}
\frac{L_\gamma({\rm on})}{L_\gamma({\rm off})} \sim \frac{\dot
\gamma({\rm on})}{\dot\gamma({\rm off})} \sim \left[{\gamma_0
\over \gamma[r_{_{\rm OB}}({\rm min})]}\right]^4\left({r_{_{\rm OB}}({\rm 
min})\over r_\gamma}\right).
\label{lum}
\end{equation} 
The maximum Lorentz factor of the primaries is (Zhang \&
Harding 2000)
\begin{equation}
\gamma_0=1.4\times 10^7P^{-1/4}(\cos\alpha)^{1/4},
\end{equation}
which is $\gamma_{0,7} \sim 1.9$ for typical values. Using Eqn (\ref{gammar}),
Eqn (\ref{lum}) and Eqn(\ref{robmin}), for $P = 0.3$,
\begin{equation} \label{Loffval}
\frac{L_\gamma({\rm on})}{L_\gamma({\rm off})} \sim 80,
\end{equation}
consistent with the typical luminosity of the Gould Belt sources.

Polar cap cascade simulations generally confirm the approximate results derived above.
A simulation run for the parameters of the Vela pulsar, using the cascade code 
developed by DH96, give the brightness and model spectra 
of $\gamma$-ray emission as a function of $\zeta-\alpha$, shown in Figures 3 and 4
respectively. The contrast between on-beam and off-beam emission will be smaller for 
longer period pulsars.   Small values of $\zeta-\alpha$ cut through the bright 
synchrotron core of the cascade, where the spectrum is a hard power law of index ~2
with a high-energy cutoff above 1 GeV due to electron-positron pair production.  
For increasing values of $\zeta-\alpha$ the 
line-of-sight falls off the main beam, views higher-altitude emission and the 
luminosity decreases.  However, the tail of high-altitude cascade curvature emission 
extends to large values of $\zeta-\alpha$ and allows the off-beam pulsars to be
visible over a large solid angle. Figure 4 shows that the off-beam emission spectrum
is much softer in the EGRET range, because the high-energy cutoff falls well below 1 GeV.
The energy of the cutoff decreases with increasing $\zeta-\alpha$, as 
$r_{_{\rm OB}}({\rm min})$ increases and critical curvature photon energy $E_0$, 
which now defines the cutoff, decreases.  The spectrum of the off-beam emission
in the EGRET band appears very soft because the $E_0$ is around 100 MeV.  Power
law fits to the off-beam model spectra between 0.1 and 1 GeV (an example is shown in
Figure 3) give photon indices in the range 2.2-3.0, consistent with what is observed 
for the Gould Belt sources. 

\vskip -0.5 truecm 
\hskip -1.0 truecm
\figureout{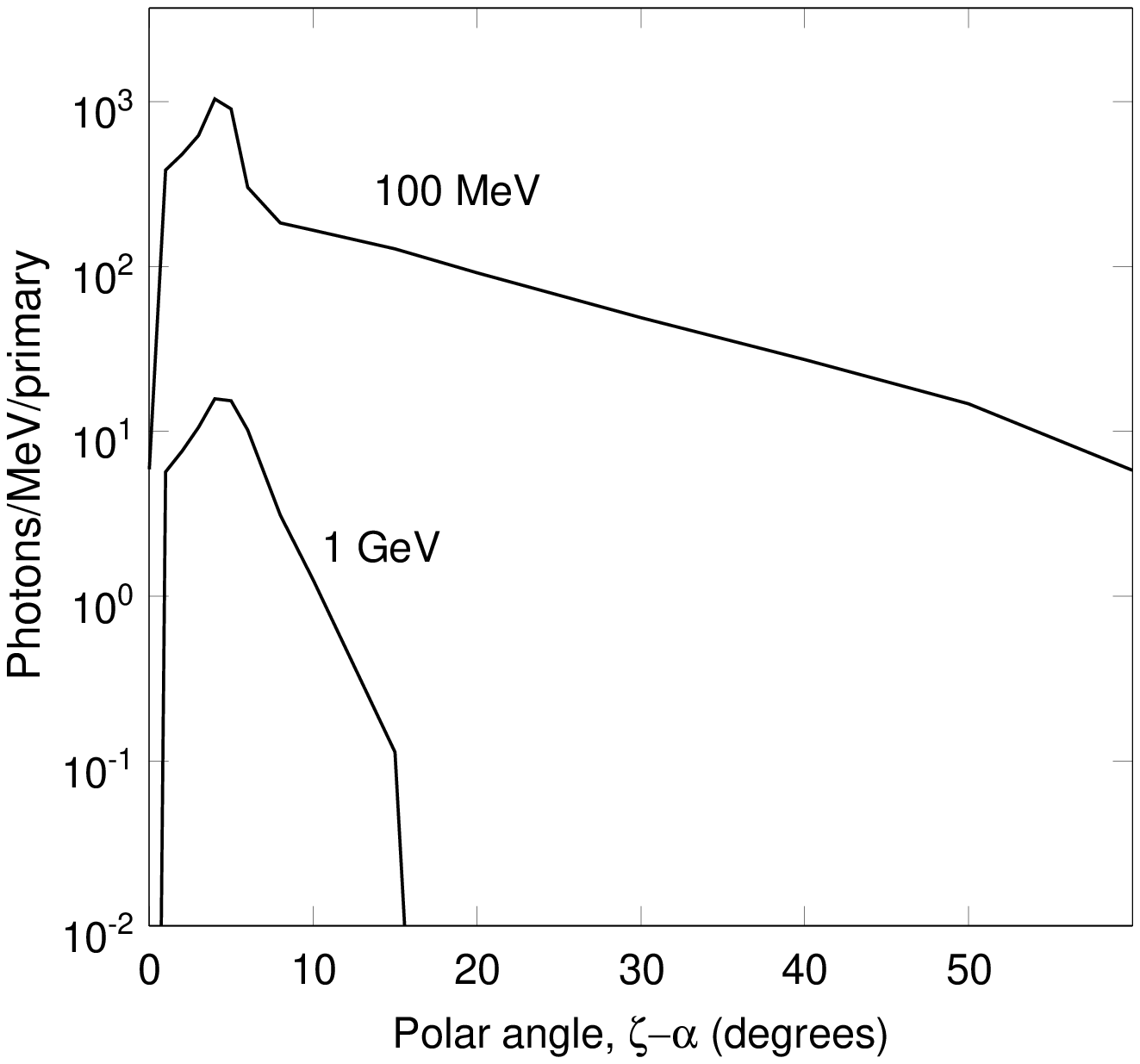}{4.4in}{0}
{Distribution of $\gamma$-ray emission at 100 MeV and 1 GeV as a function of
polar angle to the magnetic axis, $\zeta-\alpha$ for parameters of a Vela-type
pulsar (cf. Fig. 4).}
\vskip 0.5 truecm

Given the relative luminosity of off-beam pulsars calculated from Eqns (\ref{Loff}-
\ref{Loffval}), we can estimate absolute luminosity and the number of sources
detectable by EGRET and GLAST.  Using the predicted $\gamma$-ray luminosity,
$L_{\gamma}({\rm on}) \simeq 5 \times 10^{32}\,{\rm erg\,s^{-1}}$, of
a pulsar with $P = 0.3$ s and $B = 2 \times 10^{12}$ G from a polar cap cascade 
(Zhang \& Harding 2000, Eqn [60]), an average luminosity for an off-beam pulsar
would be $L_{\gamma}({\rm off}) \simeq 6 \times 10^{30}\,{\rm erg\,s^{-1}}$.  With an
out-of-plane limiting point source sensitivity of $\Phi^{EGRET} = 6 \times 10^{-8}\,{\rm 
ph\,cm^{-2}\,s^{-1}}$ (Hartman et al. 1999), we estimate that EGRET can detect
off-beam pulsars out to a limiting distance of $d_{\rm lim} \simeq 260\,{\rm pc}
(\Omega_{\rm OB})^{-1/2}$.  Using Eqn (\ref{D}) and the ratio $\theta_{_{\rm OB}}/ 
\theta_{\gamma}\sim 5$, the average solid angle for off-beam 
emission is $\Omega_{\rm OB} = 4\pi [(1-\cos\theta_{_{\rm OB}})+(\pi/2-\theta_{_{\rm OB}})\sin\theta_{_{\rm OB}}] = 3.4 \,\rm sr$.  This gives a limiting distance for
EGRET detection of off-beam pulsars as $d_{\rm lim} \simeq 140\,{\rm pc}$, about
halfway through the Gould Belt.  Assuming the supernova rate in the Belt derived
by Grenier (2000) of $75 - 95\,{\rm Myr^{-1}\,kpc^{-2}}$, there would be about
23 - 29 neutron stars of age $\lsim$ 5 Myr within 140 pc of the Sun, about 8 of
which could be detected as off-beam pulsars with a beaming factor of 
$\Omega_{\rm OB})/ 4\pi \simeq 0.27$. This estimate is consistent with the above 
relative detectability estimate derived above based on geometry. 
The GLAST out-of-plane sensitivity to point sources having a cutoff at 1 GeV is
$2 \times 10^{-8}\,{\rm ph\,cm^{-2}\,s^{-1}}$ (Digel, private comm.), which would
allow GLAST to detect off-beam pulsars out to a distance of $d_{\rm lim} \simeq 
240\,{\rm pc}$, most of the way through the Belt, and will be able to detect
pulsations of those off-beam pulsars within 150 pc.  Therefore, GLAST should be 
able to detect pulsations in all of the off-beam pulsars that EGRET detected
as point sources.

\hskip -1.0 truecm
\figureout{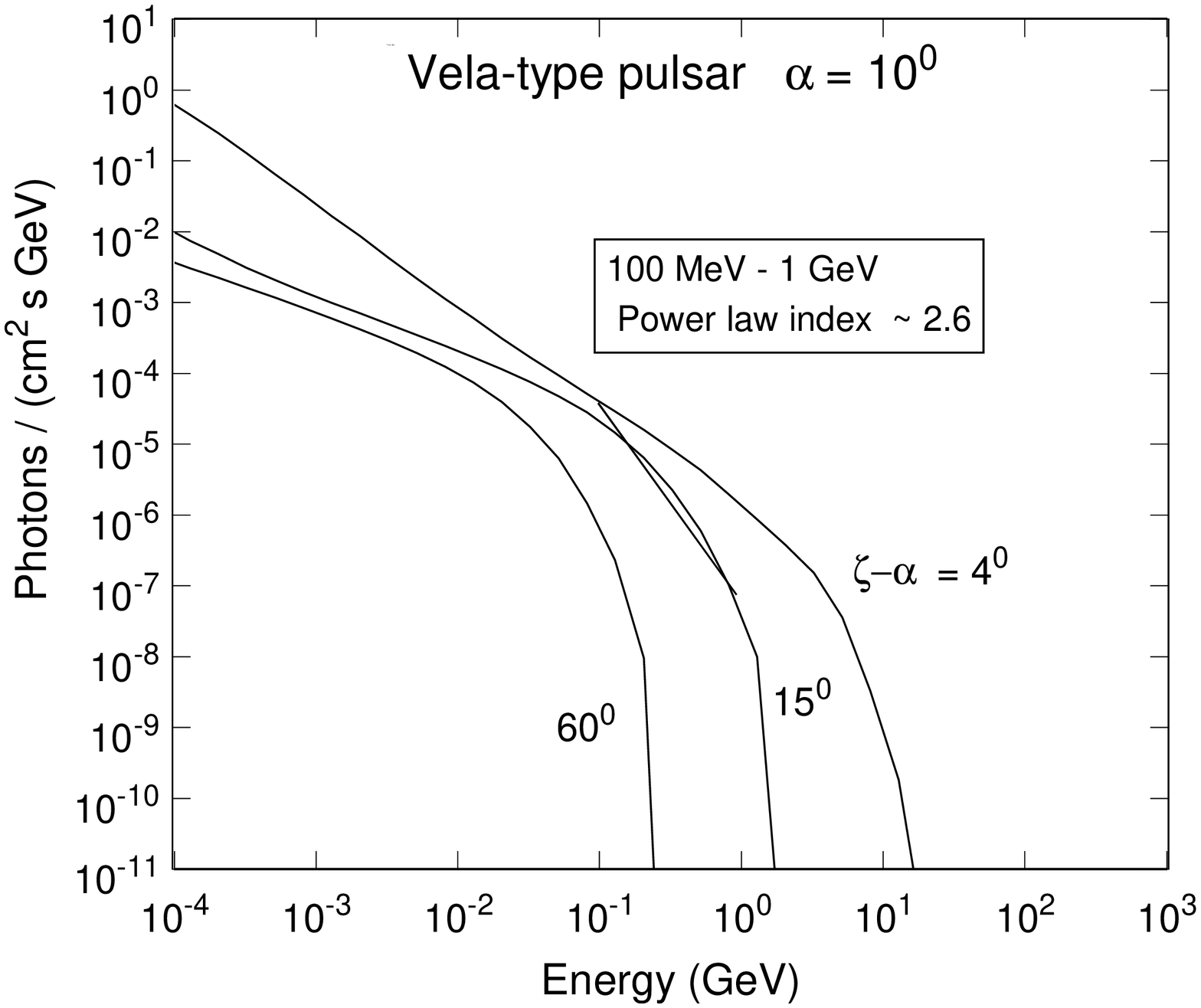}{4.4in}{0}
{Model cascade spectra for different observed 
lines of sight, $\zeta$'s. Typical Vela-type pulsar parameters, $P = 0.1$ s,
$B = 4 \times 10^{12}$ G and $\alpha = 10^0$ are 
adopted. The $\zeta-\alpha=4^0$ curve is for a typical on-beam pulsar, 
and the $\zeta-\alpha=15^0$ and $\zeta-\alpha=60^0$ 
curves are off-beam pulsar spectra. 
Notice that the off beam spectra are intrinsically fainter than the 
on-beam spectrum, and have a high-energy cut off at lower 
energies. A power law fit of the $\zeta-\alpha=15^0$ spectrum 
from 100 MeV to 1 GeV gives a spectral 
index $2.65 \pm 0.57$, consistent with the values observed from the 
unidentified ``Gould Belt sources''.
}

\section{Conclusion}

With the polar cap cascade geometry inferred from the known gamma-ray
pulsars, we find that there is a large phase space for which the line-of-sight misses
both the main gamma-ray beam and possibly the radio beam, but still cuts across 
a much broader, fainter gamma-ray beam produced by the curvature-radiation 
cooling of the primary particles.  We identify such off-beam $\gamma$-ray
pulsars as candidates for the new population of the unidentified EGRET sources 
associated with the Gould Belt.  The off-beam polar cap cascade emission exhibits
the low luminosity and soft spectrum in the EGRET band that is characteristic of 
the observed Gould Belt sources. A rough estimate indicates that the off-beam 
sources might be $\sim 4 - 5$ times more detectable than the on-beam sources 
due to the larger solid angle of the off-beam emission. A more accurate prediction 
of the fraction of off-beam $\gamma$-ray pulsars in the Gould Belt will require 
modeling of the pulsar population and luminosity selection effects. 
However, since EGRET detects all favorably oriented on-beam pulsars and a large 
fraction of off-beam pulsars in the belt, luminosity
selection effects may not be a dominant factor. 
GLAST will be able to detect pulsed emission from these
sources, and the pulse shape should be a broad single-peaked profile.

Off-beam emission is not expected in outer gap models (Yadigaroglu \& Romani 1995, 
Cheng \& Zhang 1998), which predict that most low-latitude ($|b|<5^{\rm o}$) unidentified
EGRET sources in the plane are radio-quiet, Geminga-like pulsars.  Thus, the 
radio-quiet outer-gap pulsars will be on-average as luminous in $\gamma$-rays 
as the radio-loud pulsars.  This is in contrast to our prediction that polar
cap radio-quiet pulsars (at least those detected within a few hundred pc) 
should be on average less luminous in $\gamma$-rays than 
radio-loud pulsars. 

We thank Isabelle Grenier, Matthew Baring and Seth Digel for valuable input and 
discussions.

\end{document}